\documentclass[prb,twocolumn,aps]{revtex4-1}
\usepackage{amsmath}
\usepackage{graphicx}
\usepackage{dcolumn}
\usepackage{bm}
\usepackage{hyperref}

\newcommand{\bea}{\begin{eqnarray}}
\newcommand{\eea}{\end{eqnarray}}
\newcommand{\Dv}{\bm{D}}
\newcommand{\Gv}{\bm{G}}
\newcommand{\Gc}{\mathcal{G}}
\newcommand\Gcv{{\Gc\kern-0.68em\Gc}}
\newcommand{\Sigmav}{\bm{\Sigma}}
\newcommand{\Fv}{\bm{F}}
\newcommand{\tv}{\bm{t}}
\newcommand{\kvt}{\bm{\tilde k}}
\newcommand{\dr}{\mathrm{d}}
\newcommand{\beq}{\begin{eqnarray}}
\newcommand{\eeq}{\end{eqnarray}}

\begin{document}\title{ $d$-wave superconductivity on the checkerboard Hubbard model at weak and strong coupling}
\author{\small Shiladitya Chakraborty }
\author{David S{\'e}n{\'e}chal}
\author{A.-M.S. Tremblay}
\affiliation{D{\'e}partement de Physique and RQMP, Universit\'e de Sherbrooke , Sherbrooke, QC, Canada J1K 2R1}
\date{\today}

\begin{abstract}
It has been argued that inhomogeneity generally can enhance superconductivity in the cuprate high-$T_c$ materials.
To check the validity of this claim, we study $d$-wave superconductivity on the checkerboard Hubbard model on a square lattice using the Cellular Dynamical Mean Field theory method with an exact diagonalization solver at zero temperature.
The $d$-wave order parameter is computed for various inhomogeneity levels over the entire doping range of interest in both strong and weak coupling regimes. At a given doping, the size of the $d$-wave order parameter manifests itself directly in the height of the coherence peaks and hence is an appropriate measure of the strength of superconductivity.
The weak coupling results reveal a suppression of the order parameter in the presence of inhomogeneity for small to intermediate hole dopings, while it is enhanced for large dopings.
In contrast, for strong coupling there is a monotonic decrease in the maximum amplitude of the superconducting order parameter with inhomogeneity over the entire doping range of interest.
Furthermore, at moderately high inhomogeneity, the system undergoes a first-order transition from the superconducting to the normal state in the underdoped regime.
In the overdoped regime, the change in the value of the superconducting order parameter correlates with the height of the lowest energy peak in the spectral weight of antiferromagnetic spin fluctuations, confirming the  connection between antiferromagnetic fluctuations and d-wave superconductivity found in earlier studies on the homogeneous case. Our results are benchmarked by comparisons with numerically exact results on the checkerboard Hubbard ladder.

\end {abstract}

\maketitle

\section{ Introduction}
A variety of experimental probes have revealed the existence of inhomogeneity in the electronic spectra of several families of the cuprate high-$T_c$ superconductors, which manifests itself as spatial modulation of the charge or spin density.
These manifestations of inhomogeneity include the one-dimensional stripe ordered phase observed in neutron scattering experiments \cite{tranquada} and the two-dimensional checkerboard pattern observed in optical spectroscopy measurements\cite{dordevic} and STM studies on underdoped Bi-2212  \cite{davis,vershenin,lawler} and CaNaCuOCl \cite{lupien}.
Such observations have motivated a number of theoretical scenarios involving inhomogeneity as the key ingredient in high-$T_c$ superconductivity \cite{spingap, arrigoni, caprara}.
In addition, there exist a number of theoretical results on the Hubbard model reporting an enhancement in the strength of  superconductivity in the presence of inhomogeneity \cite{kivelson1, kivelson2, kivelson3, contractor, maska, okamoto}, while other studies \cite{jarrell} find a suppression of superconductivity with inhomogeneity.
Whether inhomogeneity in the cuprates is a friend or foe of superconductivity in cuprates is therefore still an open issue.

Here we employ the Cellular Dynamical Mean-Field Theory (CDMFT) approach at zero temperature with an exact diagonalization solver to study $d$-wave superconductivity on what is commonly called (somewhat abusively) the checkerboard Hubbard model.
In contrast to previous quantum cluster based studies\cite{jarrell} using Dynamical Cluster Approximation (DCA) at finite temperature that focused entirely on the underdoped regime, we consider the entire doping range of interest from half-filling up to the extreme overdoped regime and over a wider range of inhomogeneity. In addition, we study both the weak and strong coupling regimes.

In Section \textbf{II} we introduce the checkerboard Hubbard model and describe the details of the method employed to study superconductivity. In Section \textbf{III} we present comparisons with previous accurate results \cite{kivelson4} that serve as benchmark for our approach.
We present our results in Section \textbf{IV}.
This is followed up in Section \textbf{V} by discussions and further comparisons with previous literature.
Finally, a summary of main results and final conclusions appear in Section \textbf{VI}.

\section{Method}
\begin{figure}
\includegraphics[width = 4cm]{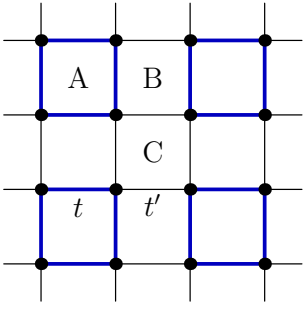}
\caption{Checkerboard lattice with hopping amplitudes $t$ (thick lines) and $t'$ on alternate bonds.
A four site plaquette can be chosen in three distinct ways namely: (1) all $t$ bonds (plaquette $A$), (2) $t$ along $y$($x$) and $t'$ along $x$($y$) (plaquette $B$) and (3) all $t'$ bonds (plaquette $C$)}
\label{check}
\end{figure}

Our starting point is the one band Hubbard model on a two-dimensional square lattice
\beq\label{hubb}
H_{\rm Hubb} =	-\sum_{i,j,\sigma} t_{ij} d^\dagger_{i\sigma}d_{j\sigma}
+ U\sum_{i} d^\dagger_{i\uparrow}d^\dagger_{i\downarrow}d_{i\downarrow}d_{i\uparrow},
\eeq
in which electrons hop among a set of lattice sites, but pay an energy cost $U$ whenever they doubly occupy the same site.
Here  $i,j$ label lattice sites, the hopping matrix elements $t_{ij}$ vanish unless $i,j$ are nearest neighbors, and $d_{i\sigma}$ annihilates an electron with spin $\sigma$ on site $i$.
As a simple toy model for inhomogeneity based on this Hamiltonian, we consider a checkerboard modulation of the nearest neighbor hopping amplitude $t_{ij}$ which varies between alternate bonds with values $t$ and $t'$  along either direction, (Figure ~\ref{check}) with $t = t'$ being the homogeneous case.

We employ CDMFT, a cluster generalization of Dynamical Mean Field Theory (DMFT)\cite{kotliar1} that allows one to reliably study d-wave superconductivity.
In CDMFT, the lattice problem is mapped to one involving a finite cluster coupled to a bath of non-interacting electrons \cite{kotliar2,maier_rmp,hettler}.
The local quantum correlations within the cluster are included exactly while longer-range correlations are treated using a mean-field approximation by writing down an effective action

\beq
\begin{split}
 S_{\rm eff} &= \displaystyle\int_0^\beta \dr\tau\, \dr\tau'\,\Psi^\dagger_d(\tau)\left[\Gcv_0^{-1}\right]\Psi_d(\tau')\nonumber \\
 &\qquad  +U\sum_\mu\displaystyle\int_0^\beta \dr\tau\, n_{\mu\uparrow}n_{\mu\downarrow},
\end{split}
\eeq
where  $\Gcv_0$ is a dynamical (time dependent) Weiss field that  describes the coupling of the cluster to the bath.
The cluster is a four-site ($2\times2$)  plaquette, which has been used extensively to study superconductivity in the Hubbard and $t-J$ models \cite{kotliar, maier, kyung}. $\Gcv_0$ contains both normal (particle-hole) as well as anomalous (particle-particle) components in order to include superconducting pairing correlations.
The Nambu spinor is defined by $\Psi^\dagger_d\equiv (d^\dagger_{1\uparrow},\cdots , d^\dagger_{4\uparrow}, d_{1\downarrow}, \cdots ,d_{4\downarrow})$ and   $\mu,\nu$ label the degrees of freedom within the cluster.

Using a starting guess for the Weiss field $\Gcv_0$, the cluster Green function $\Gv'$ is computed by solving a cluster impurity problem using a Lanczos exact diagonalization scheme, the details of which are discussed in Refs \onlinecite{kyung} or \onlinecite{senechal1}:
\beq
\Gv'(\tau,\tau')=
\begin{pmatrix}
\Gv'_\uparrow(\tau,\tau')& \Fv'(\tau,\tau')  \\
\Fv^{\prime\dagger}(\tau,\tau')&  -\Gv'_\downarrow(\tau,\tau')
\end{pmatrix},
\eeq
with $G'_{\mu\nu,\sigma}\equiv-\left\langle T d_{\mu\sigma}(\tau)d^\dagger_{\nu\sigma}(\tau')\right\rangle$ and $F'_{\mu\nu}\equiv-\left\langle T d_{\mu\uparrow}(\tau)d_{\nu\downarrow}(0)\right\rangle$, the normal and anomalous time-ordered Green functions respectively.
The cluster self-energy $\Sigmav'$ is obtained from
\beq
\Sigmav' =  \Gcv_0^{-1}-\Gv'^{-1}.
\eeq
Finally, the following self consistency condition is employed to recalculate $\Gcv_0^{-1}$ iteratively until convergence is achieved,
\beq
\Gcv_0^{-1}(i\omega_n)=
\left[\frac{N_c}{(2\pi)^2}\int \dr\kvt\,\Gv(\kvt,i\omega_n)\right]^{-1} + \Sigmav'(i\omega_n),
\eeq
with $N_c$=$4$ the cluster size and with the following definition for the superlattice Green's function
\beq\label{eq:green}
\Gv(\kvt,i\omega_n)=\left[i\omega_n + \mu -\tv(\kvt)-\Sigmav'(i\omega_n)\right]^{-1},
\eeq
where $\tv(\kvt)$ is the Fourier transform of the superlattice hopping matrix and the  momentum integral is performed over the reduced Brillouin zone of the superlattice.

The $d$-wave superconducting order parameter is defined as the expectation value of a particular pairing operator.
Since the sites on the plaquette are connected via two distinct types of links
(with hopping amplitudes $t$ and $t'$), we define correspondingly two singlet pairing operators $\hat D$ and $\hat D'$ as
\beq
\hat D = \sum_{i,j} D_{ij} d_{i\uparrow}d_{j\downarrow} \qquad
\hat D' = \sum_{i,j} D'_{ij} d_{i\uparrow}d_{j\downarrow}
\eeq
where the matrices $\Dv$ and $\Dv'$ are defined as follows: $D_{ij}=\pm 1$ on $t$-links in the $x$ and $y$ directions respectively, and likewise for  $D'_{ij}$ on $t'$ links.
The corresponding order parameters are calculated using the anomalous part $\Fv$ of the superlattice Green function
(\ref{eq:green}) as follows (for details, please consult Ref.~\onlinecite{senechal1}):
\beq\label{eq:D}
D = \frac1{(2\pi)^2}T\sum_{i\omega_n}\int\dr\kvt\,{\rm tr}\left[\Fv(\kvt,i\omega_n)\Dv(\kvt)\right]
\eeq
and likewise for $D'$.
The effective order parameter $\Psi$ is just the average $(D+D')/2$.

The hopping strengths $t$ and $t'$ are defined keeping the average bandwidth $t_0$ fixed to a constant value $t_0$ of unity as
\beq
 t  = t_0 -\Delta t  \qquad
 t' = t_0 +\Delta t
\eeq
with $\Delta t \geq 0$.
$\Delta t$  measures the degree of inhomogeneity in the system, which may be varied independently of the average bandwidth $t_0$.
We insist on the importance of keeping the latter constant when varying $\Delta t$, since varying $t_0/U$ may cause effects that are likely more important that the inhomogeneity itself.

 \begin{figure*}
\includegraphics[width=5.9cm]{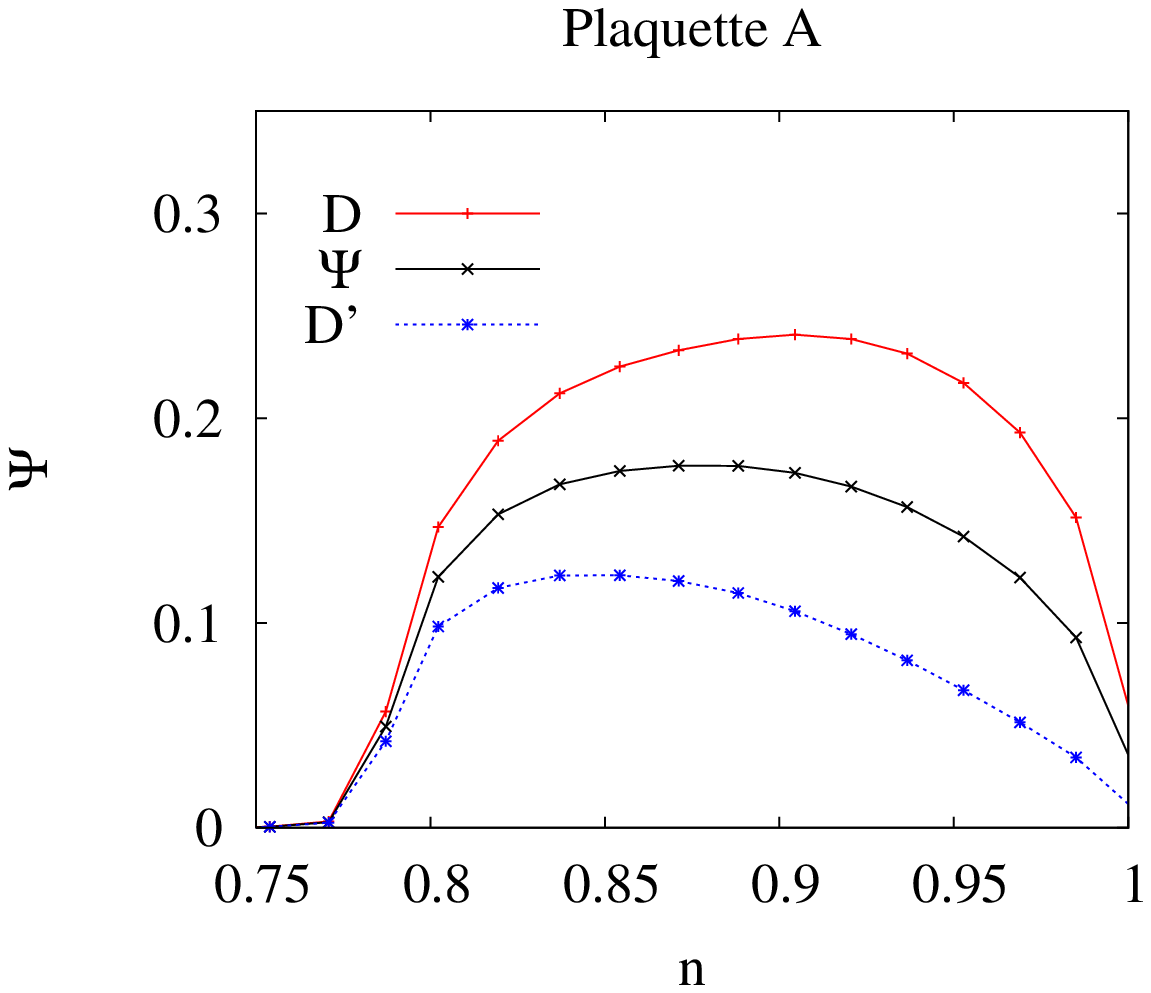}
\includegraphics[width=5.9cm]{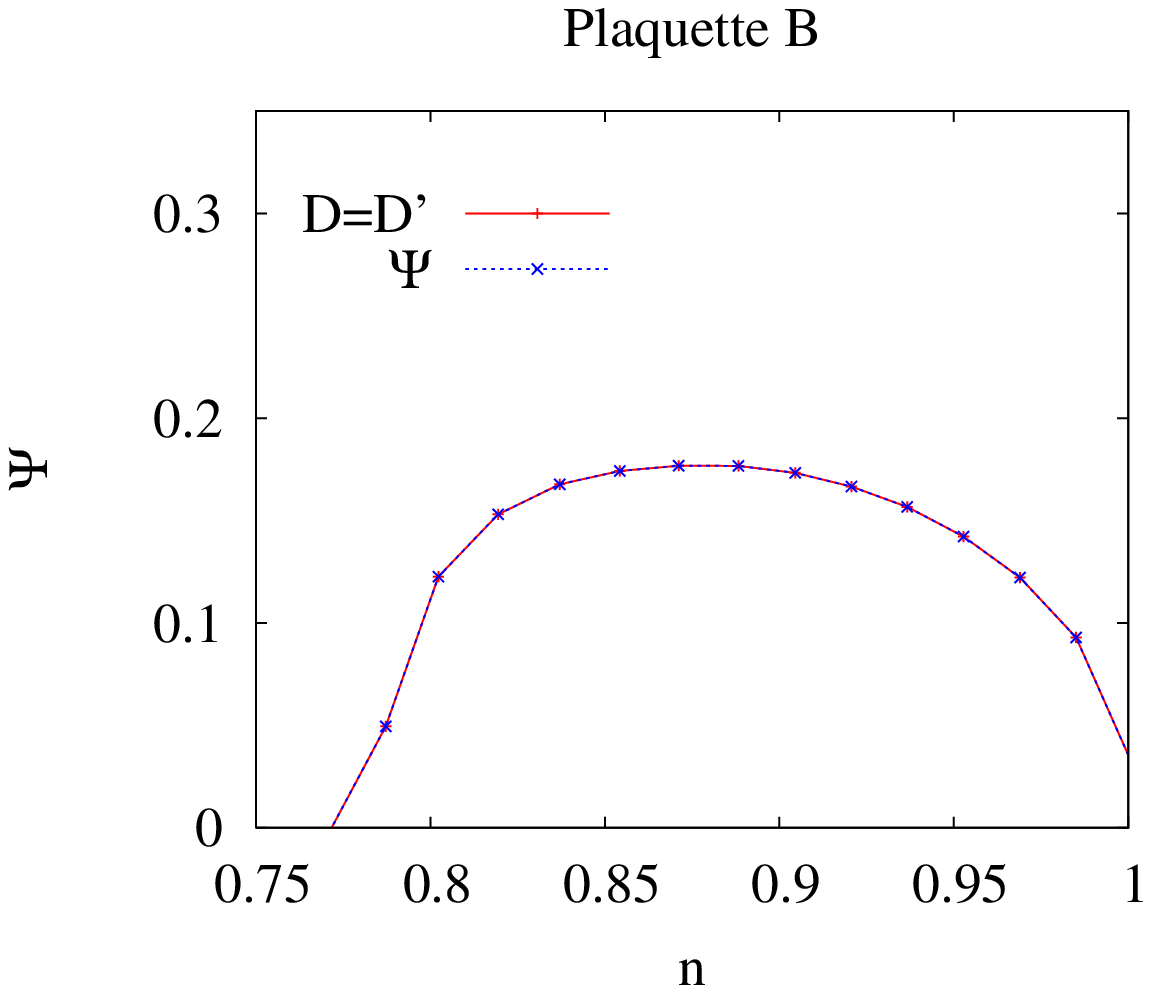}
\includegraphics[width=5.9cm]{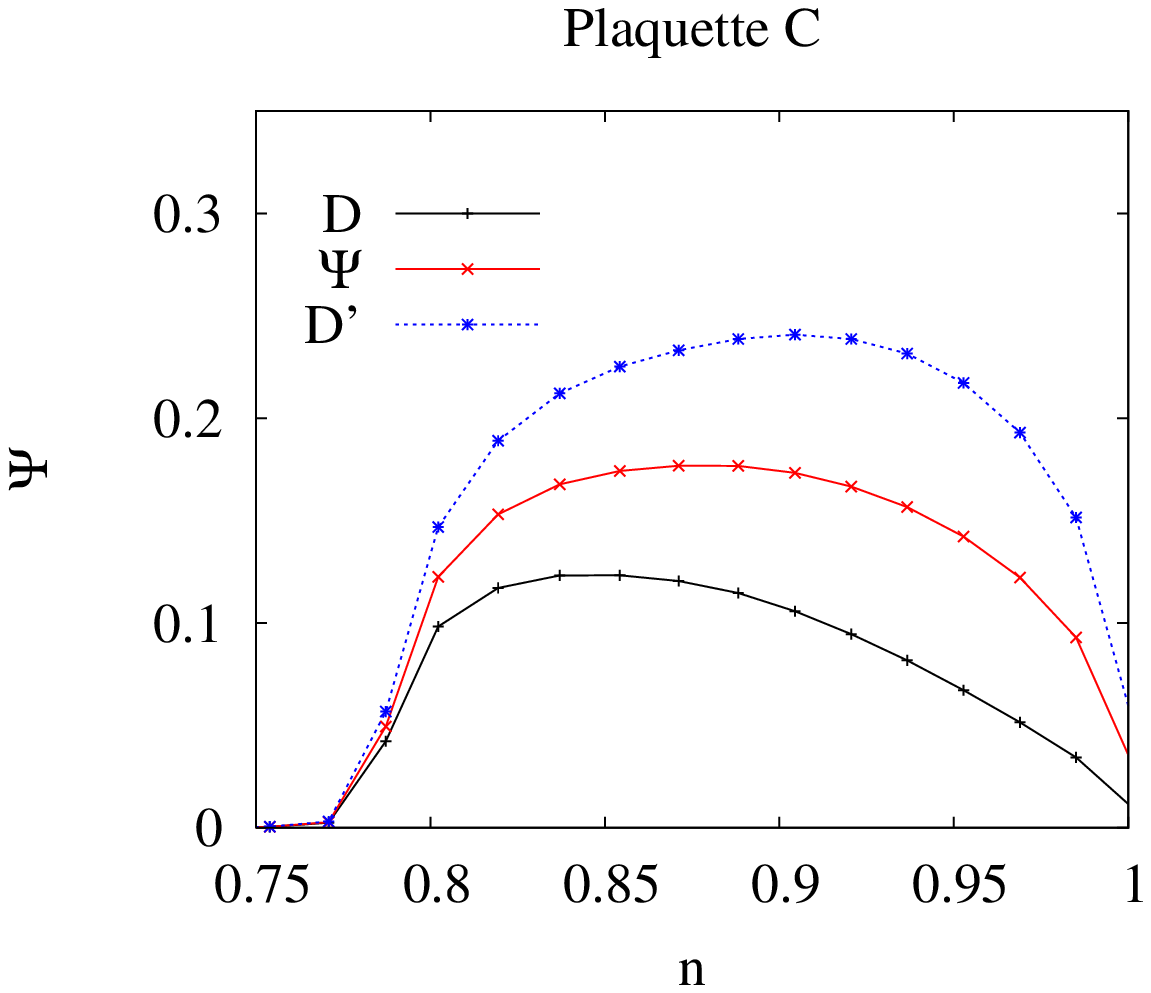}
\caption{(Color online) $d-$wave SC order parameters $D$,$D'$ for the homogeneous case ($t=t'$) for $U =8t_0$, as a function of electron filling for
plaquette $A$, $B$ and $C$. The order parameter $\Psi=(D+D')/2$ that is averaged over plaquettes $A$ and $C$ has been plotted alongside $D$,$D'$ for plaquette $B$  (middle plot) for comparison. In the latter case the curves overlap.}
\label{abc}
\end{figure*}

An immediate question arises as to the appropriate choice for the cluster.
As seen in Figure ~\ref{check}, there are three distinct ways in which a $2\times2$ plaquette might be selected, namely a plaquette with all $t$ links; one with $t$ links along $x$($y$) axis and $t'$ links along $y$($x$) axis; or one with all $t'$ links.
We label them as plaquette $A$, $B$ and $C$ respectively.
In order to select the appropriate cluster for the CDMFT calculation, one needs to determine which of the plaquettes best captures the physics of inhomogeneity.
Plaquette $B$ seems to be the natural choice under such considerations, since it includes both $t$ and $t'$ links within the cluster allowing both to be treated on the same footing within CDMFT.
In contrast, plaquette $A$($C$) treats only $t$($t'$) hoppings exactly while  $t'$($t$) are treated in a mean-field approximation via the bath degrees of freedom.

Since the link inhomogeneity expands the unit-cell by a factor of two in each direction, the four-site plaquette really constitutes a single unit cell of the model, i.e., it is not really a cluster.
As such, we are literally using single-site dynamical mean field theory instead of CDMFT, albeit with a four-band model.
Our treatment naturally collapses into a single band model (and CDMFT) in the homogeneous limit $t'=t$.

In order to further justify our choice of cluster, we compute the superconducting order parameters ($D$,$D'$) for the three clusters in the homogeneous limit ($t=t'$), where $D$ and $D'$ are expected to be identical.
We see in Figure ~\ref{abc} that, whereas the order parameters $D$ and $D'$ are identical for plaquette $B$, they are significantly different for plaquettes $A$ and $C$. To understand this result, one has to remember that even in the homogeneous case, where $t=t'$, the CDMFT lattice Green's function breaks translational symmetry, unless it is ``periodized''\cite{senechal1}.  As a consequence, the value of the order parameter on $t$ and $t'$ links will be different depending on whether they are inside or outside the plaquette. Mathematically, the self-energy matrix entering the lattice Green's function Eq.(\ref{eq:green}) is the same for all three plaquette choices when $t=t'$. However the $\Dv(\kvt)$ operators in Eq.(\ref{eq:D}) have indices that are shifted with respect to those of the hopping matrix in Eq.(\ref{eq:green}) depending on which plaquette is chosen. This also explains why, for $t=t'$, the value of $D$ on plaquette A is the same as the value of $D'$ on plaquette C while for plaquette B, $D=D'$. It should be noted however, that, as illustrated by Fig.~\ref{abc}, the mean value of the order parameter averaged over the dissimilar links is virtually identical for any choice of plaquette in the homogeneous limit.


\begin{figure}
\includegraphics[width = 8.0cm]{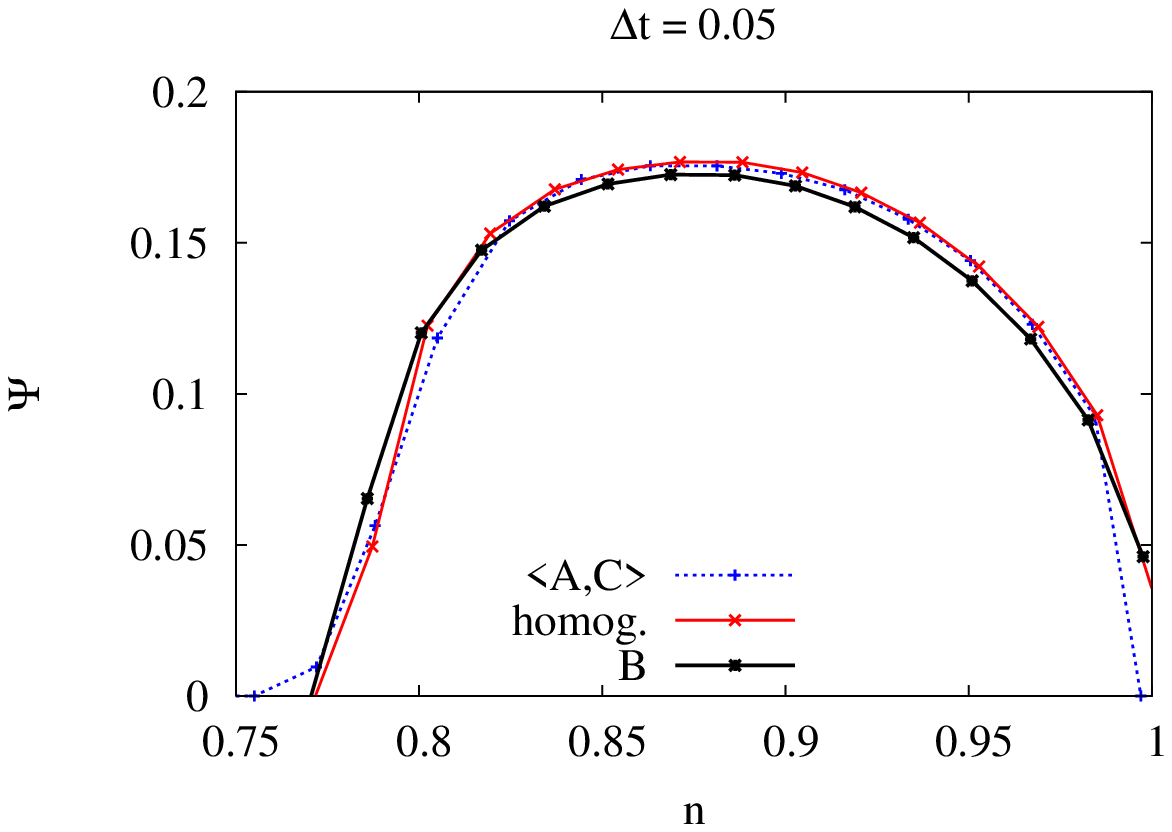}
\includegraphics[width = 8.0cm]{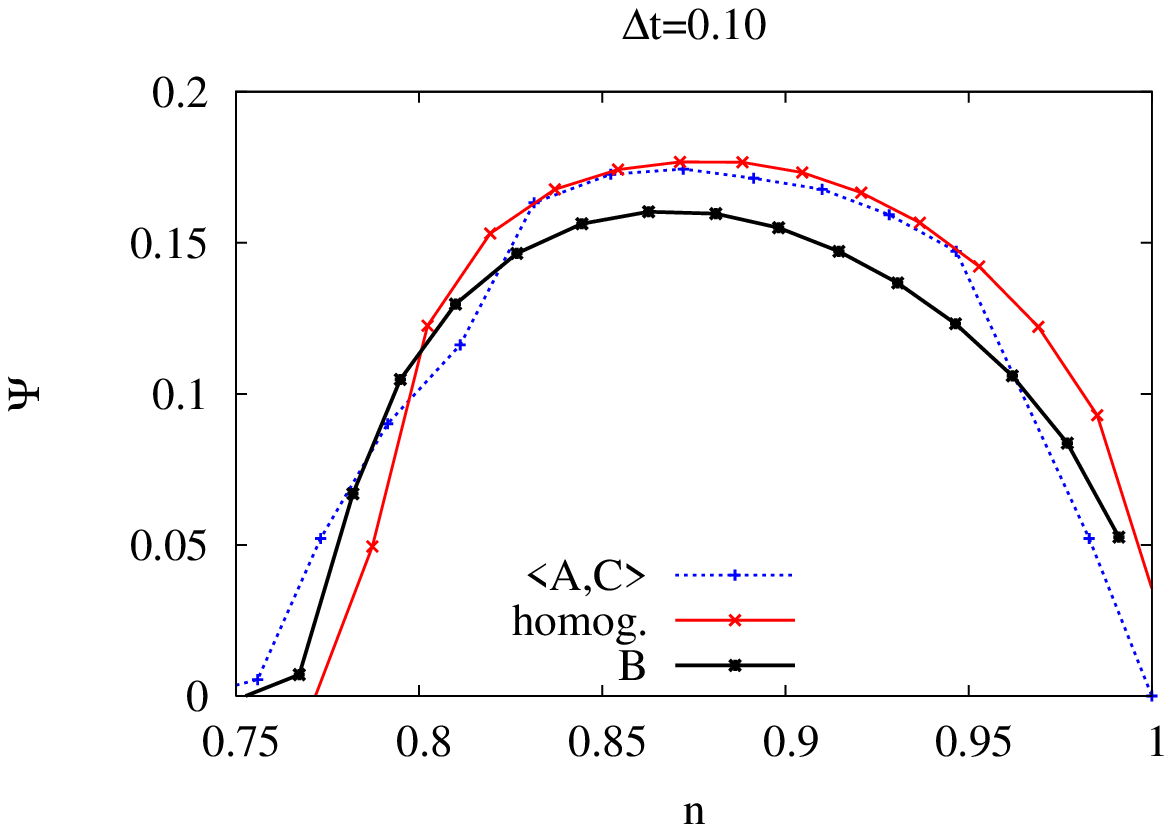}
\caption{(Color online)Average $d$-wave superconducting order parameter $\Psi$, as a function of electron filling at $U=8t_0$, $\Delta t=0.05$(top) and $\Delta t=0.10$(bottom), for plaquette $B$, the average of results for plaquettes $A$ and $C$, and the homogeneous case.}
\label{abc2}
\end{figure}

Away from the homogeneous limit, we observe  quantitative differences for different choices of plaquette.
In order to check whether or not the choice of cluster leads to a qualitative difference in our results, we compare the $d$-wave order parameter for plaquette $B$ to that of the average of plaquettes $A$ and $C$ for  moderate levels of inhomogeneity ($\Delta t = 0.05t_0$ and $0.1t_0$), as shown in Figure ~\ref{abc2}.
We find that even though the average of the superconducting order parameters for plaquettes $A$ and $C$ is larger than that for plaquette $B$, it never exceeds the corresponding values for the homogeneous case.
The results discussed in subsequent parts of this paper have been obtained using plaquette $B$ as the cluster of choice. Note that since we do not compute direction dependent quantities, the results are identical for plaquette B, whichever of the two possible $\pi/2$ related orientations we choose.

\section{Benchmark with Checkerboard Hubbard ladder}
\begin{figure}
\includegraphics[width = 5.0cm]{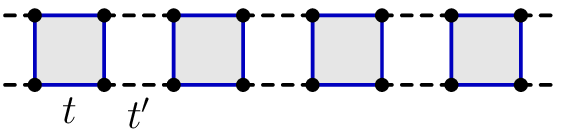}
\includegraphics[width = 8.0cm]{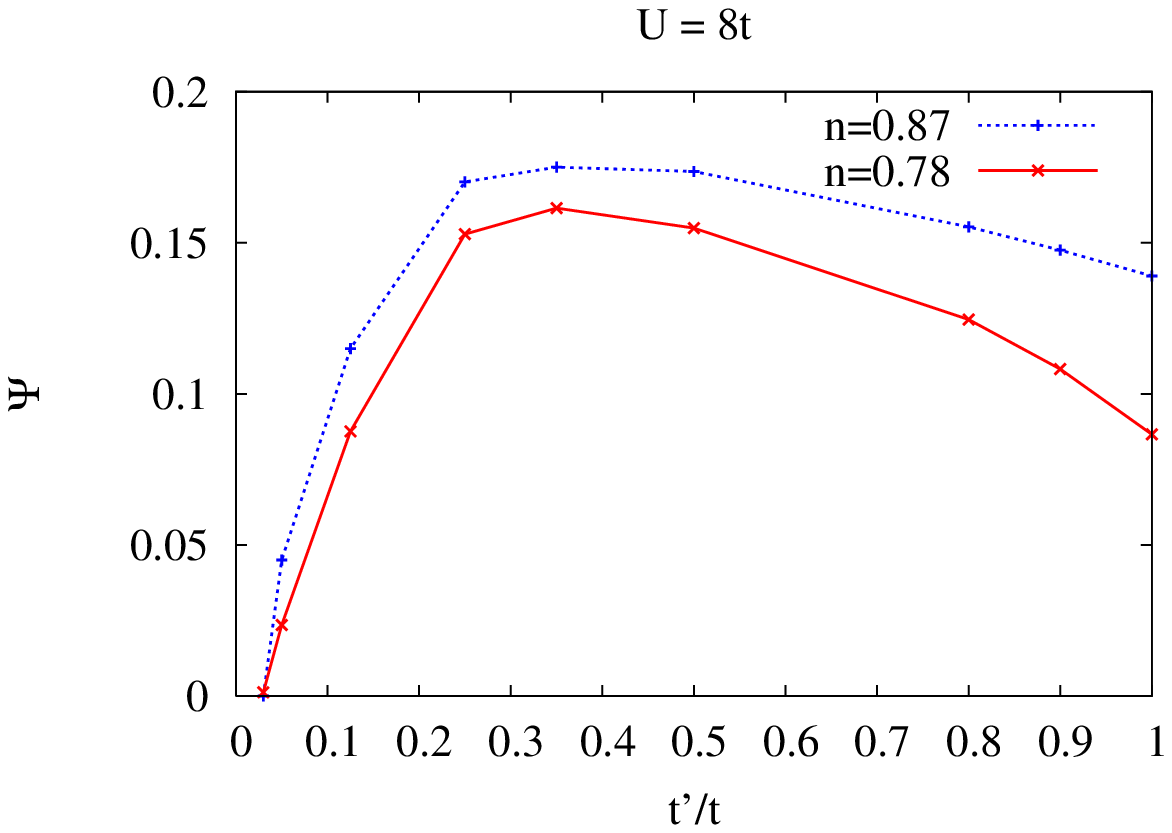}
\caption{(Color online) (Top) The checkerboard Hubbard ladder. (Bottom) $d$-wave order parameter vs $t'/t$ for two different values of electron density $n$. Note that contrary to the rest of this paper, in this subsection we work directly with the ratio $t'/t$ instead of $\Delta t$ to have a fair comparison with published results.}
\label{ladder}
\end{figure}

\begin{figure}
\includegraphics[width = 8.0cm]{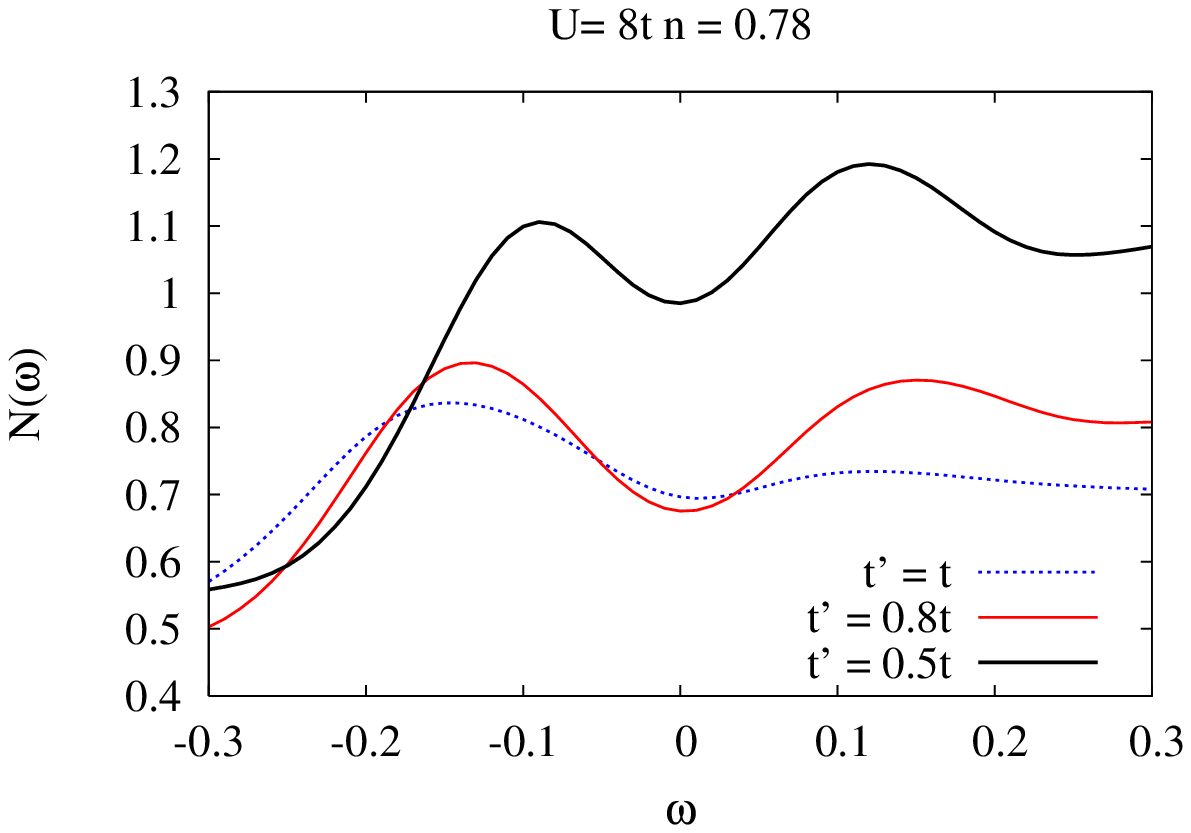}
\includegraphics[width = 8.0cm]{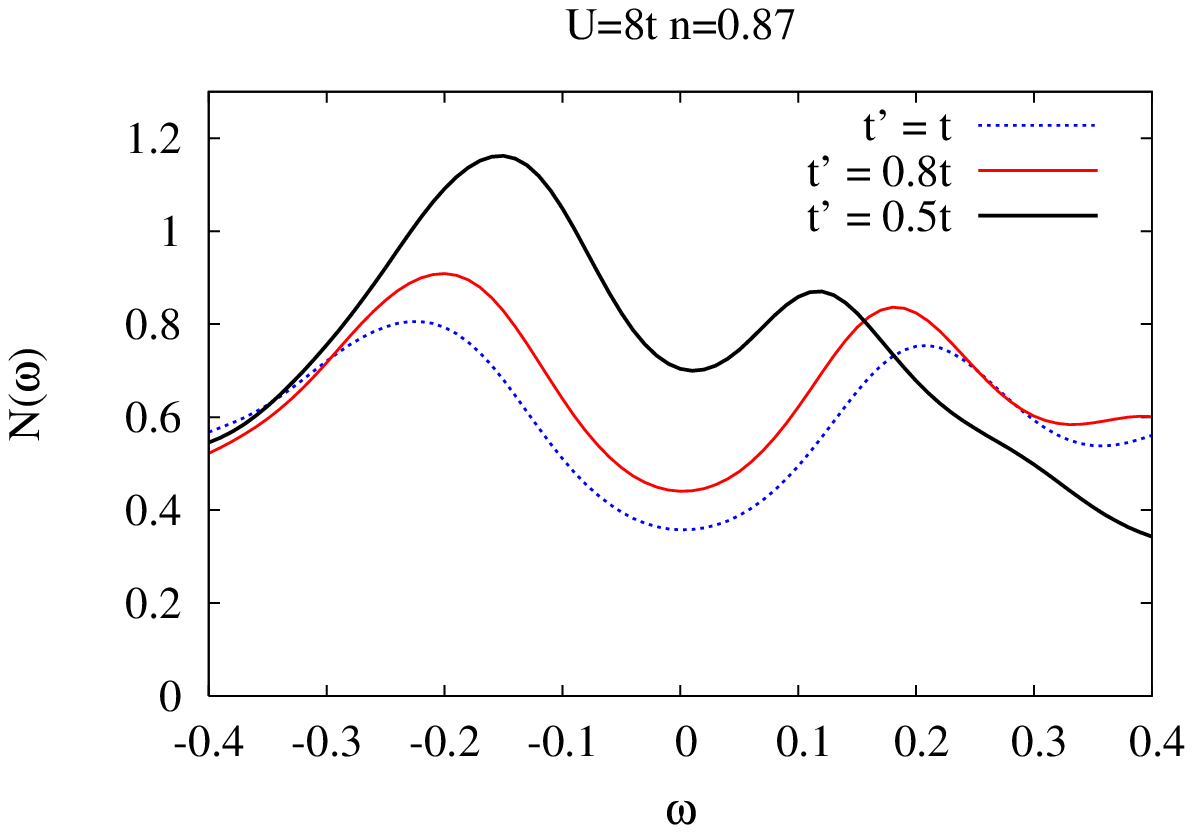}
\caption{(Color online) Single-particle density of states in the superconducting state of the checkerboard Hubbard ladder for different $t'/t$ at two values of the electron density $n$.}
\label{ladder2}
\end{figure}

To test the reliability of our CDMFT approach for inhomogeneous systems, we present results for the checkerboard Hubbard ladder, in which the Hubbard model is defined on a one-dimensional, period two array of square plaquettes consisting of $t$ links, connected by $t'$ links as illustrated in the top panel of Figure ~\ref{ladder}. This is motivated by the availability of numerically exact results for this problem that were obtained using the Density Matrix Renormalization Group (DMRG) technique \cite{kivelson4}. An enhancement in  superconductivity, as measured by the pair binding energy, was found with increasing inhomogeneity up to a moderately large value ($t'/t \approx 0.6$ for $U$=$8t$ and $n$=$0.875$), where the pair binding energy attains its  maximum value. While DMRG does not lead to long-range order on the ladder, the superconducting correlations decay algebraically. This will be mimicked by true long-range order in CDMFT, which treats long-range correlations in a mean-field way.

Our studies on the checkerboard ladder find that the dependence of superconductivity on inhomogeneity is qualitatively similar to the DMRG results. As seen in the bottom panel of Figure ~\ref{ladder}, the $d$-wave order parameter increases with inhomogeneity (decreasing $t'/t$), and attains a maximum value at a relatively larger value of inhomogeneity ($t'/t \approx 0.35$). We do not have access to the pair binding energy, as opposed to DMRG, however we can plot the density of states. The heights of the coherence peaks on each side of the energy gap in the one-particle density of states can be adopted as a measure of the strength of the superconducting correlations. These heights are indeed correlated with the magnitude of the order parameter on varying $t'/t$, as seen for two different dopings in Figure ~\ref{ladder2}. Such behavior is in stark contrast to that of the two-dimensional checkerboard Hubbard model, where, as we shall discuss below, there is no  non-zero optimal value of  inhomogeneity that favors superconductivity. Regardless of the contrasting observations in the two systems, this exercise serves to strengthen our claim of the validity of our CDMFT results for the two-dimensional system.
\begin{figure}
\includegraphics[width = 8.0cm]{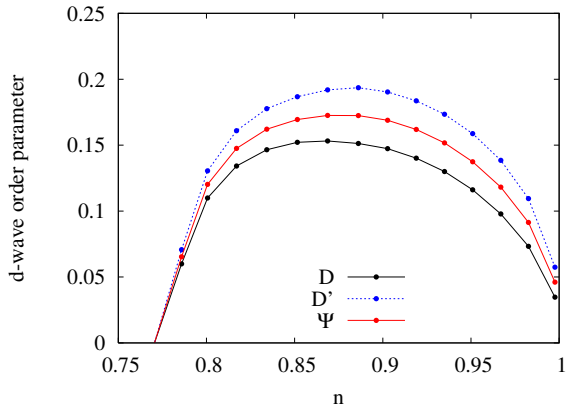}
\caption{(Color online) $d-$wave SC order parameters $D$,$D'$ and their average $\Psi$ as a function of electron filling $n$ for $U$ = $8t_0$ and $\Delta t = 0.05$ computed with plaquette B. }
\label{ddp}
\end{figure}

\section{Results}
This section is divided into three parts.
We first describe the results of computation of the $d$-wave superconducting order parameter for strong and weak coupling, without allowing for antiferromagnetic long-range order. In addition, we show that in the superconducting state at a given doping, the heights of the peaks of the density of states lying on either side of the Fermi energy (across the energy gap) correlate with the magnitude of the order parameter. We then show that the correlation found previously\cite{senechal} between the low energy peak in the imaginary part of the spin susceptibility and the d-wave order parameter is still preserved in the inhomogeneous case.

\subsection{Superconducting order parameter}
Figure ~\ref{ddp} shows $D$, $D'$ and $\Psi$ plotted as a function of the electron density $n$ for $U = 8t_0$ and $\Delta t = 0.05$.
The strength of $d-$wave superconductivity over the entire doping range is larger across a link with larger hopping amplitude ($t'$).
This is consistent with other studies \cite{kyung} which find that in the strong-coupling limit $\Psi$ scales roughly with $J = 4t^2/U$, the nearest-neighbor spin super-exchange coupling (at $U = 8t_0$ we are indeed entering the strong-coupling regime).
Quantitative differences aside, the plots of $D$ and $D'$ otherwise look qualitatively very similar, arising from the fact that  $t$ and $t'$ are both treated on the same footing within the plaquette thereby allowing us to systematically isolate the physics of inhomogeneity from that of  varying the effective bandwidth.

\begin{figure}
\includegraphics[width = 8.3cm]{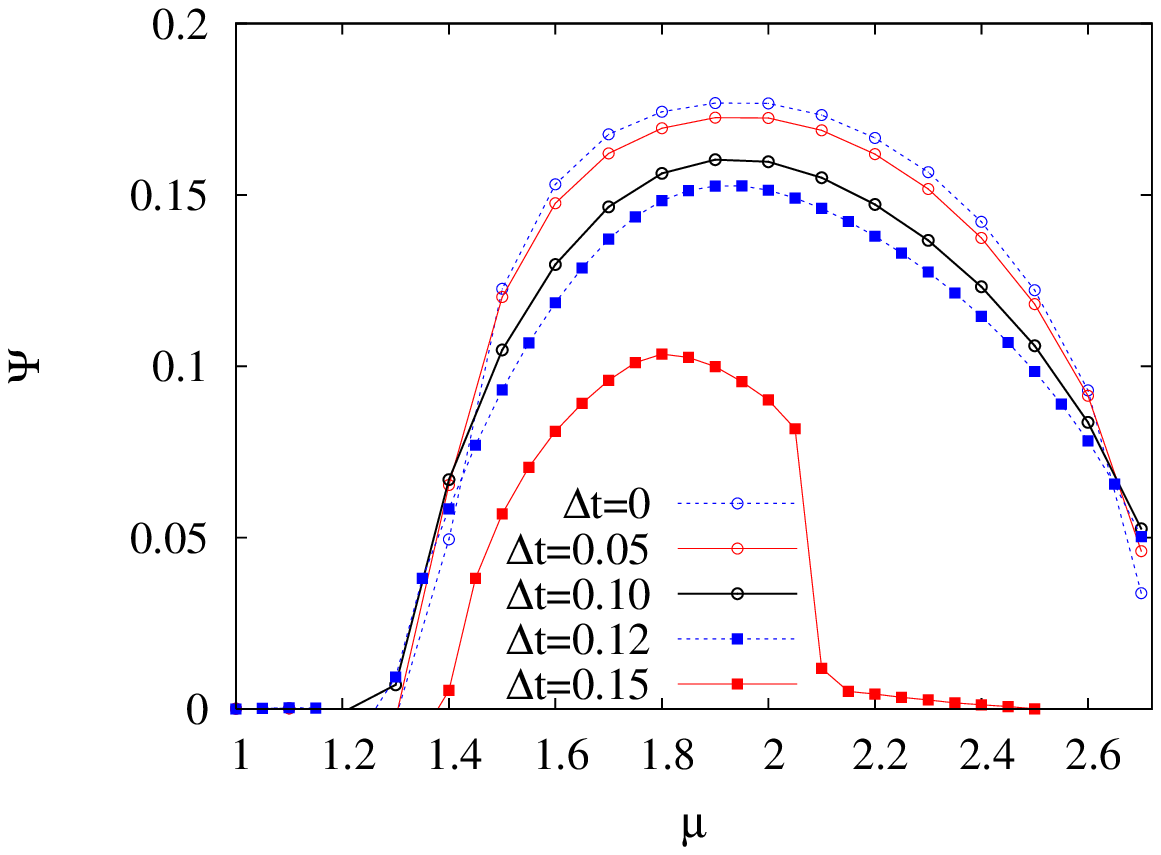}
\includegraphics[width=8.3cm]{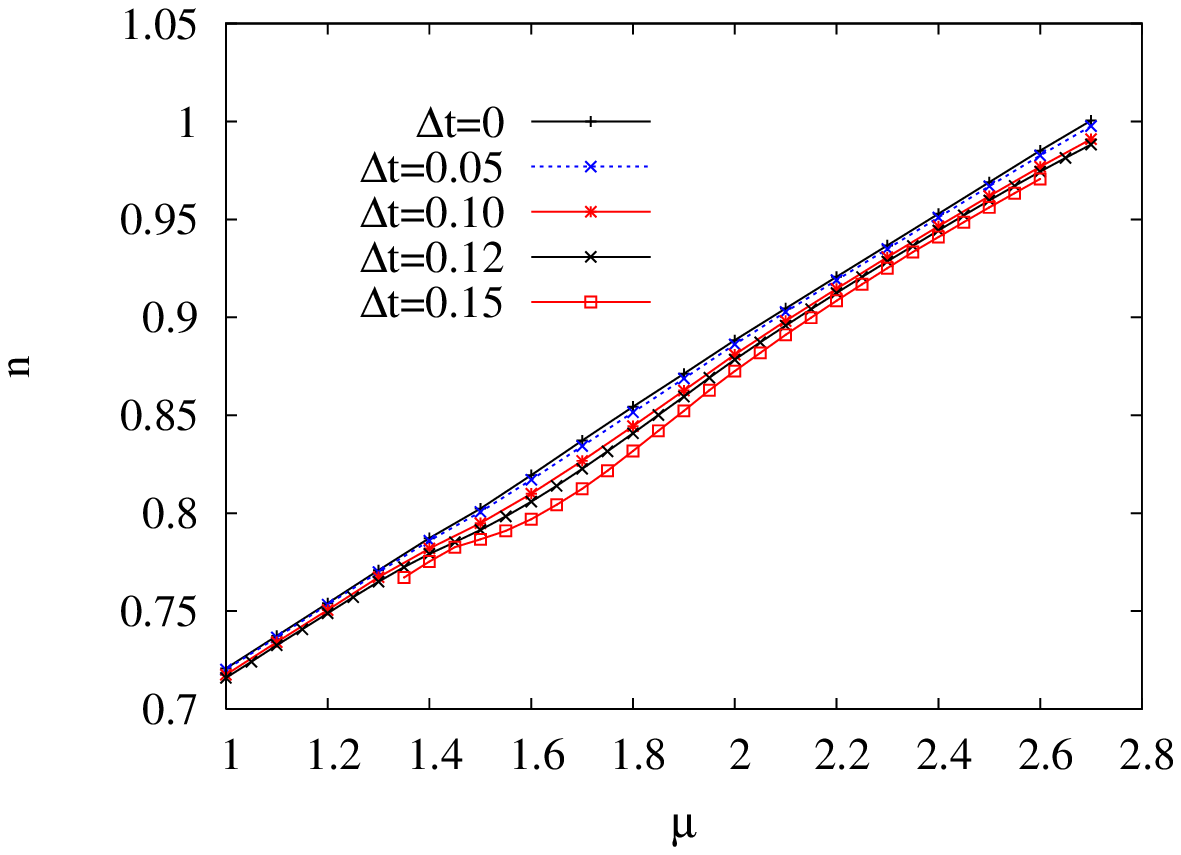}
\caption{(Color online)(Top)$d$-wave superconducting order parameter $\Psi=(D+D')/2$ as a function of chemical potential for various values of inhomogeneity $\Delta t$ for plaquette B. (Bottom) Dependence of the lattice electron density on the chemical potential for the values of $\Delta t$ that appear in the above plot.}
\label{inhomo}
\end{figure}

\begin{figure}
\includegraphics[width = 8.0cm]{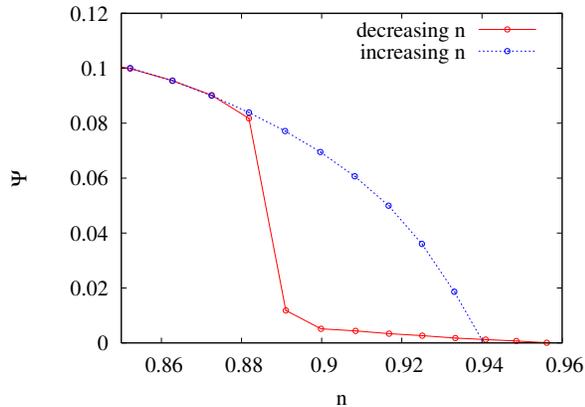}
\caption{(Color online) Average $d$-wave superconducting order parameter as a function of electron filling $n$ for $\Delta t$= 0.15, showing hysteretic behavior as the system is tuned between the normal and the superconducting states in the underdoped regime. The blue (dashed) curve corresponds to increasing $n$, starting from a superconducting initial state close to optimal doping, while the  red (bold) curve corresponds to reducing $n$, starting from a normal initial state close to half-filling. The value of the order parameter in the filling range of (0.89-0.94) is dependent on the initial state being normal or superconducting.}
\label{hysterisis}
\end{figure}

The results for the superconducting order parameter for several values of inhomogeneity $\Delta t$, displayed in the top panel of Figure~\ref{inhomo}, exhibit two interesting features.
We find that the strength of $d$-wave superconductivity decreases monotonically as a function of $\Delta t$ over the entire doping range over which superconductivity exists. This stands in contrast to results obtained by studies on finite clusters \cite{kivelson1, kivelson2, kivelson3, contractor} where the pair binding energy was found to be maximized for moderately high levels of inhomogeneity at low hole dopings.
However, our findings are in qualitative agreement with those of Doluweera \textit{et. al.} \cite{jarrell}, in which DCA was used to study the checkerboard Hubbard model and the superconducting transition temperature $T_c$ was found to fall monotonically as a function of inhomogeneity in the underdoped regime.
In addition, we observe  a first-order superconducting to normal transition in the underdoped regime beyond a moderately large level of inhomogeneity ($\Delta t \geq0.15t_0$).
The existence of the first-order transition is confirmed by the observation of hysteretic behavior in the order parameter depending on the initial state being normal (small doping) or superconducting (larger doping) (Figure ~\ref{hysterisis}), as the system is tuned across a superconducting to normal transition.
Finally, as inhomogeneity is increased further to $\Delta t \geq 0.16t_0$ (corresponding to $t/t'\leq 0.72$, we find that superconductivity is completely destroyed for all dopings.

\begin{figure}
\includegraphics[width = 8.0cm]{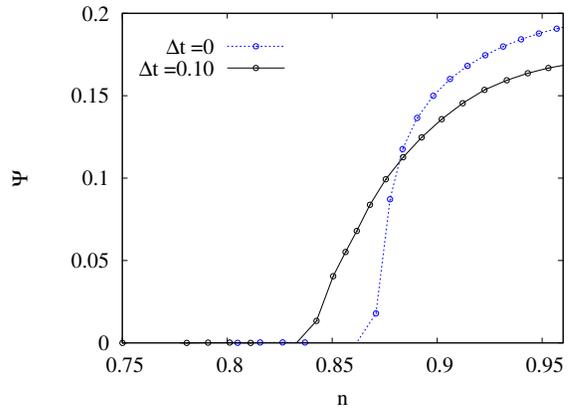}
\caption{(Color online) Average $d$-wave superconducting order parameter $\Psi=(D+D')/2$, as a function of electron filling at $U=4t_0$, for homogeneous limit ($\Delta t=0$) and $\Delta t =0.1t_0$ for plaquette B.}
\label{u4}
\end{figure}

So far we have focused on strong coupling  ($U=8t_0$), where the system is a Mott insulator at half-filling.
It has been shown \cite{kyung} that in the strong-coupling regime,  proximity to the Mott insulating state leads to the suppression of superconducting order parameter close to half-filling.
In contrast, the behavior is very different for weak coupling, where the Mott transition is absent, and no suppression is observed in the $d$-wave order parameter in the underdoped regime unless antiferromagnetic long-range order is allowed\cite{kyung}.
Results for $U=4t_0$ in the presence of inhomogeneity are shown in Figure ~\ref{u4}.
We find that the superconducting order parameter in the inhomogeneous case ($\Delta t = 0.10t_0$)  is suppressed compared to the homogeneous case except except for large dopings, where the superconducting order parameter is larger in the inhomogeneous case. The maximum value of the order parameter which occurs at half-filling in both cases, is however larger in the homogeneous case. We must emphasize here that for weak coupling, the CDMFT method is not completely reliable as longer range antiferromagnetic correlations, which are important at weak coupling close to half-filling, are not adequately captured by this technique. Nevertheless, these results may serve to demonstrate the qualitative difference between the  weak coupling and strong coupling results.

\subsection{Density of states}
The principal disadvantage of using the $d$-wave order parameter as a measure of superconductivity is that the order parameter, though easily computed within CDMFT, is not an  experimentally measurable quantity, in contrast to the superconducting energy gap or $T_c$. Within the BCS theory, the gap is given by the order parameter times the effective interaction\cite{tinkham}. Regardless of the validity of BCS theory for superconductivity in the Hubbard model, if one assumes that such a relation holds approximately true, a knowledge of the effective interaction would be required in addition to the order parameter to estimate the gap. In particular, in order to accurately estimate the gap as a function of $\Delta t$, one would have to determine the dependence of the effective interaction on inhomogeneity.

\begin{figure}
\includegraphics[width = 8.0cm]{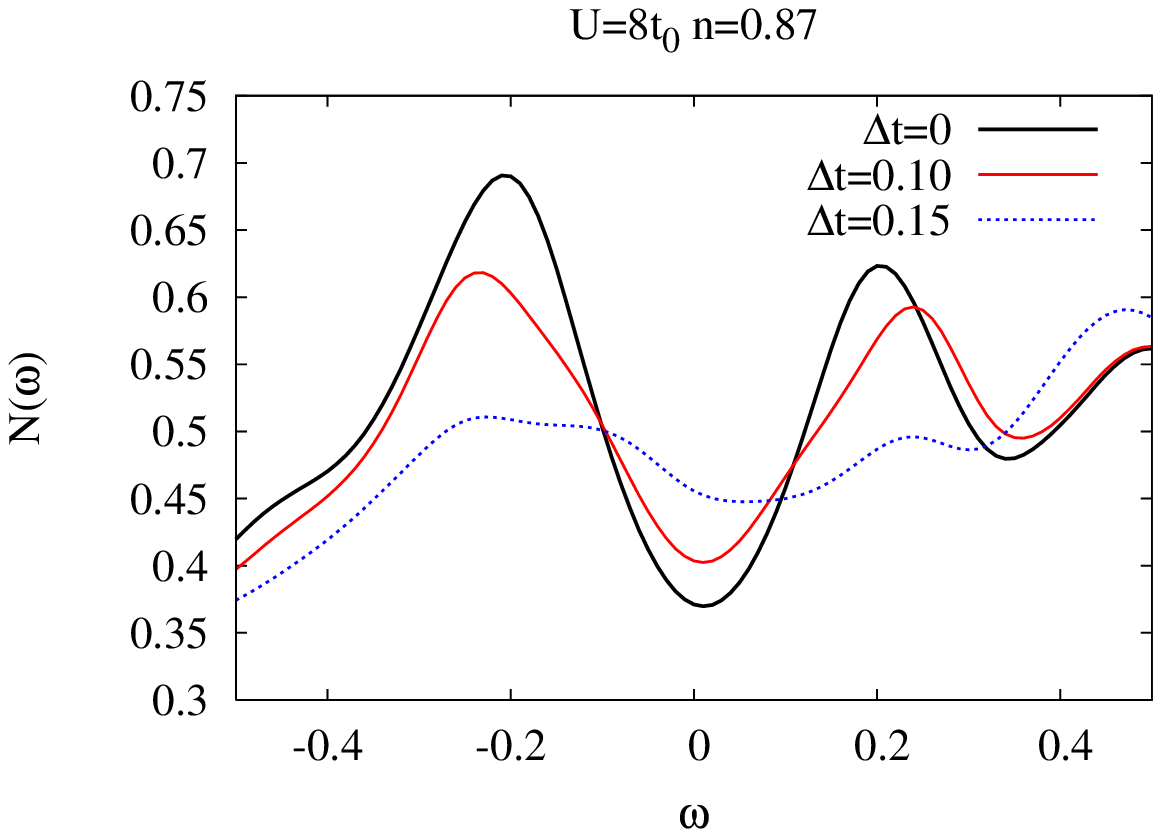}
\includegraphics[width = 8.0cm]{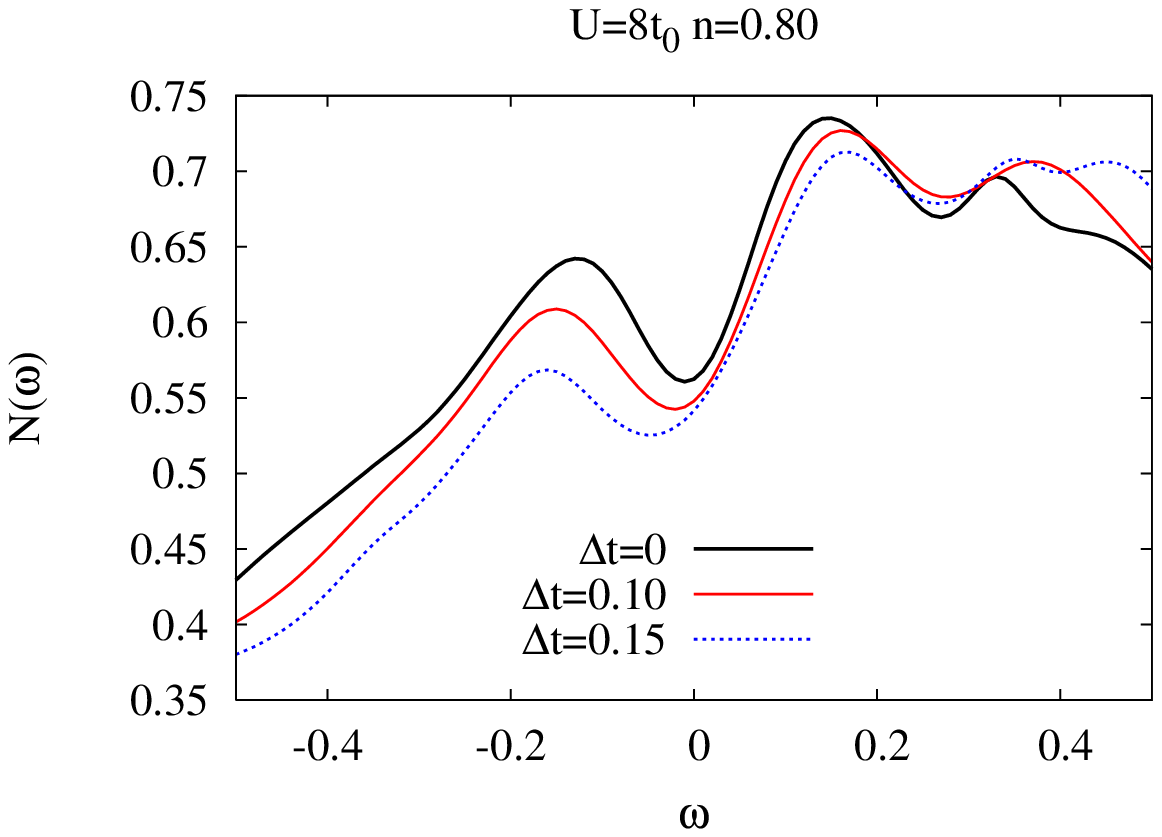}
\caption{(Color online) Single-particle density of states for different values of $\Delta t$ at two values of the electron density $n$ for plaquette B.}
\label{DOS1}
\end{figure}

A more straightforward and physically meaningful way to estimate the strength of superconducting correlations is to compute the single-particle density of states, which is directly measured in tunnelling experiments\cite{tinkham}, for varying $\Delta t$ (Figure ~\ref{DOS1}). The density of states features a gap around the Fermi energy as seen in earlier studies\cite{kyung}. However, the magnitude of the energy gap does not change appreciably with $\Delta t$ on the scale of the Lorentzian broadening $\eta$= $0.1t_0$, used to compute the density of states, which makes it unsuitable to probe reliably the variation of superconductivity with inhomogeneity. Alternatively, one might consider the quasiparticle peak heights in the density of states. The quasiparticle spectrum of BCS superconductors is characterized by peaks on either side of the energy gap, and the height of the peaks is a measure of the coherence of the quasiparticle excitations, and may therefore be used to gauge the strength of coherence in the superconducting state for varying $\Delta t$ at a fixed doping. As seen in Figure ~\ref{DOS1}, for various $\Delta t$ the heights of the peaks vary concomitantly with the magnitude of the $d$-wave order parameter(Figure~\ref{inhomo}) in both underdoped and overdoped regimes, thereby providing an experimentally measurable probe of superconductivity whose behavior is consistent with that of the order parameter.

\subsection{Spin susceptibility}
\begin{figure}
\includegraphics[width = 8.0cm]{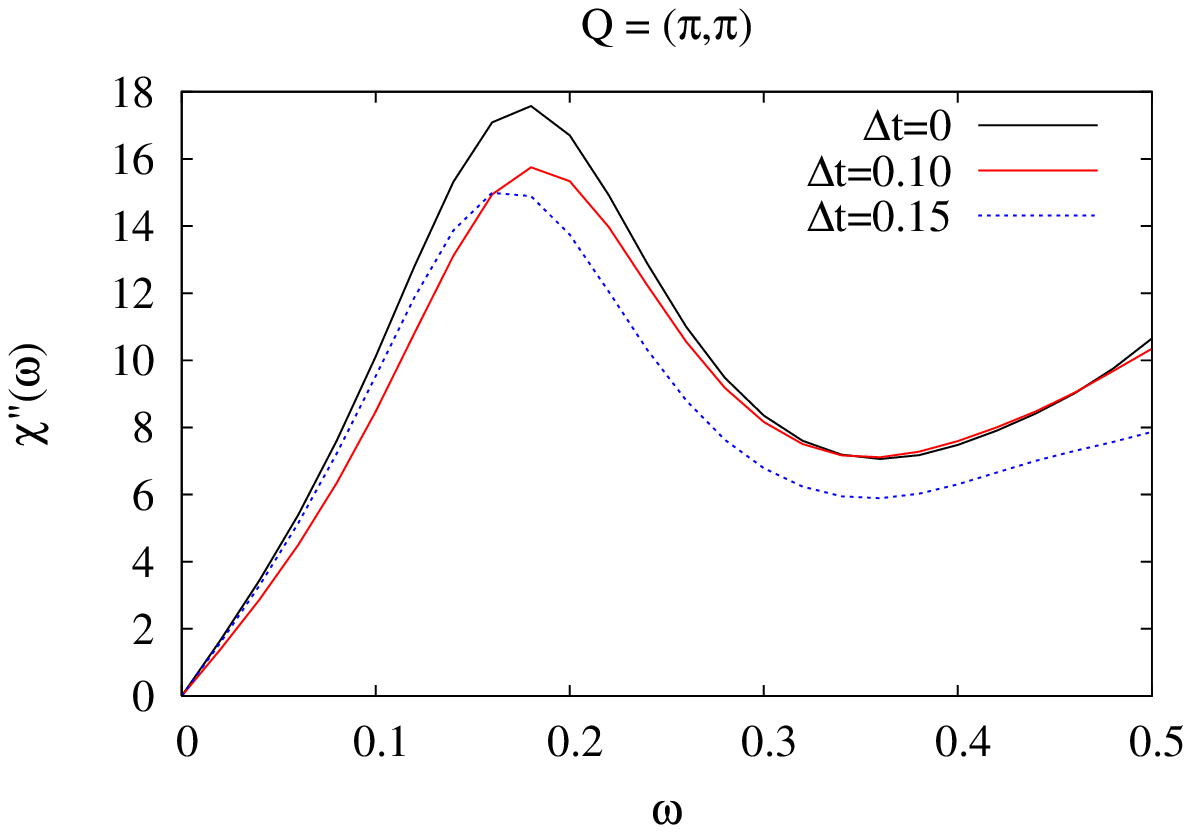}
\includegraphics[width = 8.0cm]{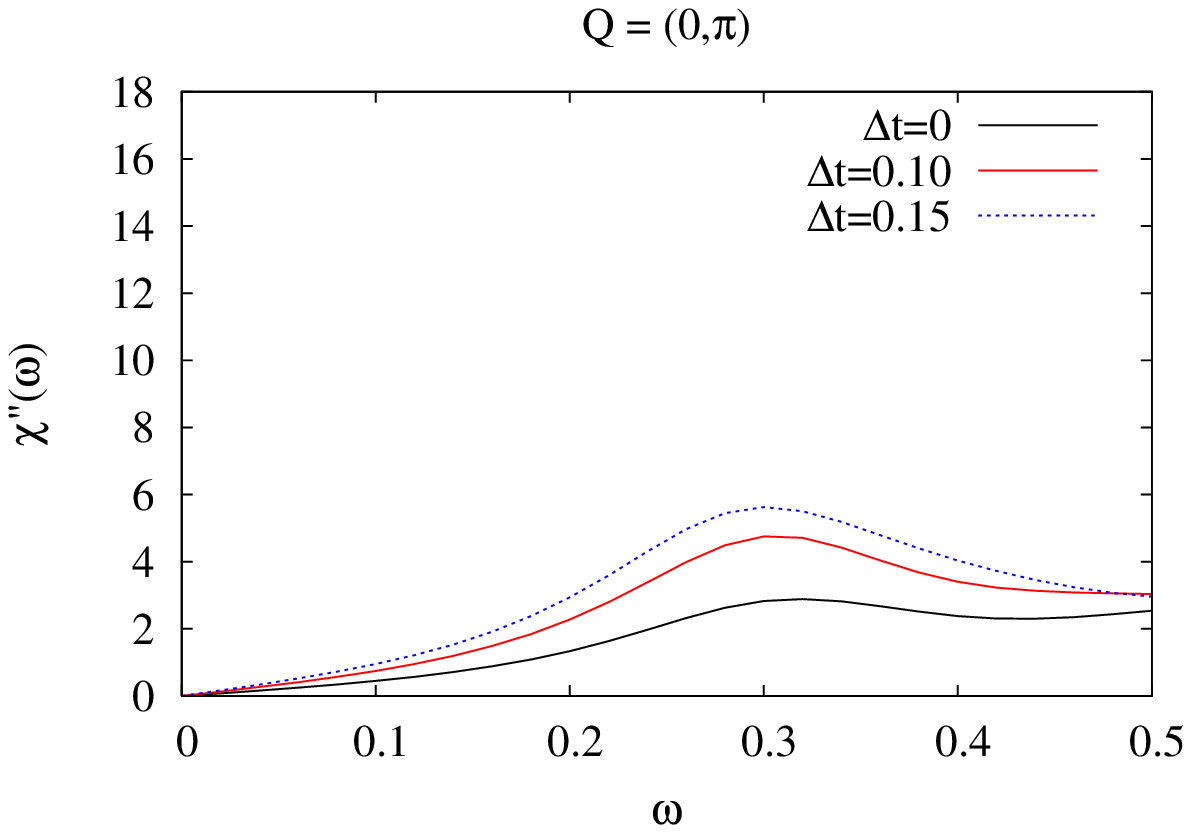}
\caption{(Color online) Imaginary part of the cluster spin susceptibility at $\textbf{Q}$=$(\pi, \pi)$ (top) and $\textbf{Q}$=$(0,\pi)$ (bottom), for three different $\Delta t$ values, with $U$=$8t_0$ and $n$=0.84.}
\label{ss}
\end{figure}

\begin{figure}
\includegraphics[width = 8.0cm]{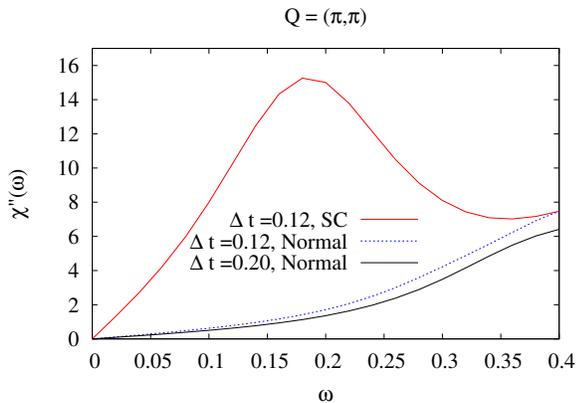}
\caption{(Color online) Imaginary part of the cluster spin susceptibility at $\textbf{Q}$=$(\pi, \pi)$ in the superconducting ($\Delta t$=0.12, $n$=0.84), and normal state ($\Delta t$=0.12, $n$=0.76 and $\Delta t$=0.20, $n$=0.77) for $U$=$8t_0$.}
\label{ss2}
\end{figure}

The top panel in Figure ~\ref{ss} displays the imaginary part of the cluster spin susceptibility $\chi''(\omega)$ in the overdoped regime ($n$= $0.84$), for $U$= $8t_0$ at $\textbf{Q}$ = $(\pi,\pi)$ for three different values of $\Delta t$.
The strength of the low energy peak, whose connection with superconductivity  was confirmed by theoretical studies \cite{senechal,maier2} on the overdoped homogeneous Hubbard model and neutron scattering experiments \cite{wakamoto1,wakamoto2,dahm} on LSCO samples, falls with increasing inhomogeneity, concomitant with the behavior of the superconducting order parameter. In contrast with the  homogeneous case, the fall in the susceptibility peak with inhomogeneity is slower than that of $d$-wave superconductivity.  Nevertheless, we find that the association between superconductivity and the low energy peak in the spin susceptibility at $\textbf{Q}$ =$(\pi,\pi)$ in the overdoped regime is preserved in the presence of inhomogeneity as well. As seen in Figure ~\ref{ss2}, the low energy peak in the spin susceptibility is present in the superconducting state, while it disappears in the normal state ($\Psi = 0$), which further corroborates the  connection between antiferromagnetic fluctuations and $d$-wave superconductivity found previously in the homogeneous Hubbard model.

Another interesting feature in the spin susceptibility is the enhancement of the  $\textbf{Q}$ =$(0,\pi)$ component of $\chi''(\omega)$ with inhomogeneity as seen in the bottom panel in Figure ~\ref{ss}, which may be an indication of development of  spin fluctuations competing with the predominant antiferromagnetic fluctuations in the system. One must however be careful in interpreting the above results, since the cluster which is inherently anisotropic, favors such spin fluctuations, and  does not necessarily reflect on the nature of the long-range spin correlations.
The strength of the low energy peak (around $\omega$=$0.3$)  however remains significantly smaller than that at $\textbf{Q}$ = $(\pi,\pi)$, indicating that antiferromagnetic correlations, although weakened by inhomogeneity, continue to dominate the physics of the two-dimensional Hubbard model in the presence of moderate checkerboard-type inhomogeneity.

\section{Discussion}
Our studies indicate that in the strong coupling regime, which is considered relevant for the cuprates, checkerboard-type inhomogeneity on the square lattice Hubbard model is detrimental to $d$-wave superconductivity over the entire doping range of interest.  Superconductivity is completely destroyed beyond a moderately large inhomogeneity level.
This is a strikingly different conclusion from what was reported in some recent works that studied the highly underdoped regime of the checkerboard Hubbard model on finite clusters using exact diagonalization\cite{kivelson1,kivelson2,kivelson3}, and contractor renormalization \cite{contractor} methods, where the pair binding energy was found to be maximum at moderate levels of inhomogeneity at low hole concentrations.

The difference between these results and those of the present study may be attributed to several reasons.
Firstly, CDMFT is substantially better equipped  in capturing the physics of the extended lattice, compared to finite clusters which  are expected to have significant finite-size effects.
Therefore, interpreting the results obtained on finite systems and using them to predict the nature of superconductivity in the extended system must be undertaken cautiously.
At least CDMFT captures some of the physics of the infinite lattice in a mean-field way, even though correlations are taken into account exactly only on short length scales.
Secondly, previous studies computed the pair binding energy to quantify superconductivity which, in contrast to the $d$-wave order parameter considered in our study, is not a measure of  superconducting phase coherence in the system.
Furthermore, our results are in qualitative agreement with those of Doluweera \textit{et al.}\cite{jarrell}, who used DCA on a four-site plaquette to study superconductivity on the checkerboard Hubbard model at finite temperature.
Their study reported a monotonic suppression of $T_c$ as a function of inhomogeneity.
In contrast, our work focused on the $d$-wave order parameter at $T=0$ as a measure of superconducting strength, providing an alternative approach to this problem.
It is noteworthy that in contrast to the aforementioned work where a single plaquette configuration was studied  with uniform hopping within the cluster (corresponding to plaquette $A$($C$) in our work depending on $t'$ being larger (smaller) that $t$), we have verified the dependence of results on the choice of cluster.
It is not surprising, however, that our results are qualitatively similar, since DCA and CDMFT are both self-consistent cluster methods that effectively capture the short-range correlations of the system, and whose results become exact in the limit of infinite cluster size.

There is a qualitative difference between the strong and weak coupling results in the extreme overdoped regime where there is an enhancement in superconductivity at weak coupling that is not observed at strong coupling.
It is also worth mentioning here that our investigations include the overdoped regime of the superconducting phase of the  checkerboard Hubbard model, which has not been considered in any of the previously mentioned studies.

In the strong coupling case, the gradual suppression of the order parameter with inhomogeneity, followed by a first-order transition to the normal state at moderately large inhomogeneity in the underdoped regime has not been noticed before.
This result should be interpreted in light of a recent study that demonstrates that a first-order metal-metal transition lies beneath the superconducting dome \cite{Sordi} and that this transition is directly linked to the Mott transition.
The superconducting phase of the unusual metal found close to half-filling is more sensitive to inhomogeneity.
This can be verified by looking at Fig.~\ref{inhomo} at fixed filling as a function of inhomogeneity: the figure clearly suggests that a first order transition to the normal state occurs.

Finally, our results support the connection between antiferromagnetic fluctuations and superconductivity found previously for the homogeneous case. \cite{Maier3,senechal} Indeed, the correlation between the magnitude of the superconducting order parameter and the height of the  first peak in $\chi''$ in the overdoped regime \cite{senechal} remains valid for the inhomogeneous case studied here, although the fall of the susceptibility peak is slower than that of the order parameter.
It is entirely plausible therefore that a suppression in the superconductivity is tied to the of weakening of antiferromagnetic correlations in the presence of a checkerboard-type inhomogeneity in nearest-neighbor hopping.

\section{Conclusion}
We benchmarked our approach with numerically exact DMRG results on the checkerboard Hubbard ladder \cite{kivelson4}. The quantities that can be obtained in CDMFT and in DMRG are different but qualitatively both approaches show similar results, namely that there exists an optimal inhomogeneity for superconductivity on a ladder.

Despite suggestions that this result is general, namely that there is always an optimal inhomogeneity for superconductivity \cite{kivelson1, kivelson2, kivelson3, contractor, maska, okamoto, Chakravarty, Martin, Aryanpour, Loh, Arrigoni}, our study demonstrates instead that this statement may not be valid for d-wave superconductivity in the two-dimensional Hubbard model in the presence of arbitrary types of inhomogeneity.
Indeed, previous DCA results at finite temperature \cite{jarrell} and our CDMFT study at zero temperature both find that when inhomogeneity on the checkerboard Hubbard model is in the nearest-neighbor hopping, either the maximum $T_c$ \cite{jarrell} or the maximum value of the superconducting order parameter at zero temperature cannot exceed that of the homogeneous system. The size of the order parameter at a given doping can be taken as a measure of the strength of superconductivity, since we have shown that it manifests itself directly in the height of the coherence peaks in the density of states.
Note that since one finds, with CDMFT, that \textit{site} inhomogeneity on the checkerboard lattice does lead to an optimal inhomogeneity for superconductivity,\cite{okamoto} quantum cluster methods do not have intrinsic limitations that prohibit finding enhanced superconductivity in the presence of inhomogeneity.

For the model of interest, we have explored a larger doping range than previous studies as well as both weak and strong coupling regimes.
In the weak coupling case, as a result of inhomogeneity, superconductivity is suppressed in the underdoped and enhanced in the overdoped regime, which however does not surpass the maximum possible value of the order parameter in the homogeneous case. In the strong coupling case, our results can be summarized by the following observations a) a monotonic suppression of the superconducting order parameter for all dopings  and b) at given inhomogeneity, a first order transition between normal and superconducting state at finite doping on the underdoped side, with the superconducting dome disappearing suddenly on further increasing the inhomogeneity.

Further research could look into the effect of next-nearest-neighbor hopping or of mixed types of inhomogeneity, including both hopping and site energies.
This would help again to verify the generality of the connection confirmed here between antiferromagnetic fluctuations and superconductivity. \cite{Maier3,senechal} One should also explore more closely the relationship between the domain where a first order transition is induced by inhomogeneity at strong coupling in the underdoped regime and the domain where a first order transition between two metals was found recently for the homogeneous case. \cite{Sordi} The latter phenomenon was clearly linked to the Mott transition.

\begin{acknowledgments}
The authors would like to acknowledge Steven Kivelson for discussions. This work was partially supported by NSERC and by the Tier I Canada Research Chair Program (A.-M. S. T.).
Computational resources were provided by CFI, MELS, the RQCHP, and Compute Canada.
\end{acknowledgments}


\end{document}